# LES MÉTHODES DE RÉSOLUTION DE TYPE « LATTICE BOLTZMANN » SONT-ELLES UTILISABLES POUR SIMULER LES JETS PLASMAS SOUFFLÉS ATMOSPHÉRIQUES UTILISÉS EN PROJECTION ?

Ridha **DJEBALI** [a], Nicolas **CALVÉ** [b], Bernard **PATEYRON** [b], Mohammed **El GANAOUI** [b,*], Habib **SAMMOUDA** [a]

[a] LETTM, Faculté des Sciences de Tunis, Université de Tunis El Manar 2, Tunisie
[b] SPCTS, Faculté des Sciences et Techniques, Université de Limoges, France

**RÉSUMÉ**

Dans cette étude une nouvelle approche numérique, à savoir la méthode de Boltzmann sur réseau couplé au modèle de turbulence de Smagorinsky, est employée pour simuler et modéliser le comportement de jets plasma axisymétriques et turbulents. Un mélange de gaz plasmagène argon-azote (N2-Ar 62.5% vol.) est utilisé. La variation des paramètres de diffusion (viscosité et diffusivité) en fonction de la température est bien prise en compte. La qualité des résultats montre une grande efficacité du présent modèle par comparaison aux modèles de codes GENMIX et Jets&Poudres pour la dynamique des jets plasma.

*Mots Clés : Jets plasma, méthode de Boltzmann, modèle axisymétrique*

## NOMENCLATURE

**Symboles :**

$f_k, g_k$   fonctions de distribution de la densité et de la température
$f_k^{eq}, g_k^{eq}$ distributions d'équilibre de $f_k$ et $g_k$
**u**   vitesse $=(u_x, u_y)$, m.s$^{-1}$
**x**   noeud du réseau en coordonnées (x,y)
$e_k$   vitesse discrète du réseau
$\Delta t$   pas de temps, s
$\Delta x$   pas d'espace, m
$C_s$   constante de Smagorinsky
T   température, K

Lettres grecques :
$\rho$   Densité du fluide, kg.m$^{-3}$
$\upsilon$   viscosité cinématique, m$^2$.s$^{-1}$
$\alpha$   diffusivité thermique, m$^2$.s$^{-1}$
$\tau_\nu, \tau_\alpha$   temps de relaxation, s
$\Delta$   épaisseur du filtre, m

Indices / Exposants:
eq   équilibre
k   direction de la vitesse discrète
tot   total
en   entrée

## 1. INTRODUCTION

Les procédés de projection plasma jouent un rôle économiquement important et ouvrent des perspectives dans le développement de nouvelles technologies. Plusieurs travaux expérimentaux et numériques sont menés sur ce thème dans le but d'atteindre des performances élevées (en traitements de surfaces et revêtement) en raison des contraintes économiques et afin de comprendre les transferts complexes et les propriétés des transports couplés de chaleur et de masse. C'est que les distributions de température du jet plasma et des différents champs d'écoulement, au cœur des flux, affectent les trajectoires des particules et leurs histoires thermiques, et régissent donc la qualité de projection thermique (dépôt obtenu).

Les travaux de modélisation (de jets plasma) disponibles sont presque tous fondés sur l'hypothèse d'écoulement stationnaire en sens de moyenne temporelle. Cependant, il a été démontré [1, 2, 3] que le jet plasma est instationnaire. La méthode de Boltzmann sur réseau (LBM) est fondamentalement adaptée à simuler les écoulements de gaz, qui présentent un processus collisionnel.

* auteur correspondant
*Adresse électronique :* ganaoui@unilim.fr

Bien que l'intérêt à l'approche de la resolution LBM soit croissant, ces modèles sont principalement limités aux systèmes de coordonnées cartésiennes. Quelques modèles axisymétriques ont été développés récemment. Le modèle axisymétrique de Jian Zhou [4] a été employé avec succès par R. Djebali et al. [5] pour simuler un jet plasma d'argon employant un modèle turbulence LES-LBM. Dans l'étude présentée, le comportement de jet plasma est étudié pour un mélange de gaz, à savoir le N2-Ar 62.5% vol.[6, 7] Le modèle axisymétrique est couplé au modèle de turbulence standard de Smagorinsky. Les résultats de ce modèle sont validés par comparaison à différentes références et montrent un niveau de prédiction élevé.

## 2. MODELISATION

Les équations qui régissent l'écoulement, écrites sous des hypothèses adéquates (voir [8]), sont résolues en utilisant une formulation thermique de la méthode de Boltzmann sur réseau dans la limite de l'incompressibilité [9, 10] et en tenant compte du caractère turbulent et des paramètres de viscosité, conduction et diffusion fonctions de la température [11, 12]. La méthode est décrite comme suit.

### 2.1. MODELE AXISYMETRIQUE THERMIQUE

Les équations discrétisées du modèle proposé sont données pour une double population par:

$$\begin{cases} f_k(\mathbf{x} + \Delta\mathbf{x}, t + \Delta t) - f_k(\mathbf{x}, t) = \\ \quad -\frac{[f_k(\mathbf{x},t) - f_k^{eq}(\mathbf{x},t)]}{\tau_\nu} + \Delta t F_1 + \frac{\Delta t}{6} \mathbf{e}_{ki} F_{2i}, \quad k = 0,8 \\ g_k(\mathbf{x} + \Delta\mathbf{x}, t + \Delta t) - g_k(\mathbf{x}, t) = \\ \quad -\frac{[g_k(\mathbf{x},t) - g_k^{eq}(\mathbf{x},t)]}{\tau_\alpha} + \Delta t S, \quad k = 1,4 \end{cases} \quad (1)$$

Où S, $F_1$ et $F_{2i}$ sont des fonctions d'espace dérivant de la formulation axisymétrique. Les temps de relaxation $\tau_\nu$ et $\tau_\alpha$ sont liés à la viscosité cinématique et la diffusivité thermique par $\nu = \frac{\tau_\nu - 0,5}{3}\frac{\Delta x^2}{\Delta t}$ et $\alpha = \frac{\tau_\alpha - 0,5}{2}\frac{\Delta x^2}{\Delta t}$, davantage d'informations sur le modèle sont données en [3].

Les variables macroscopiques sont calculées utilisant les moments d'ordres zéro et un comme suit:

$$\begin{cases} \rho(\mathbf{x},t) = \sum_{k=0,8} f_k \\ \mathbf{u}(\mathbf{x},t) = \frac{1}{\rho} \sum_{k=0,8} \mathbf{e}_k f_k \\ T(\mathbf{x},t) = \sum_{k=1,4} g_k \end{cases} \quad (2)$$

### 2.2. MODELE DE TURBULENCE

En modélisation LBM-LES, la viscosité est ajustée localement par addition de la viscosité turbulente à la viscosité moléculaire. Pour le réseau D2Q9 de la méthode de Boltzmann, la viscosité effective obéit à l'équation suivante:

$$\nu_{tot} = \frac{\tau_{\nu\text{-tot}} - 0,5}{3} = \nu + \nu_t = \nu + (C_s \Delta)^2 |\bar{S}_{ij}| \quad (3)$$

Des calculs intermédiaires conduisent à une équation du second degré, qui donne:

$$\tau_{\nu-tot}(\mathbf{x},t) = \left(\tau_\nu + \sqrt{\tau_\nu^2 + 18(C_s\Delta)^2 |Q_{ij}|/\rho(\mathbf{x},t)}\right)/2 \quad (4)$$

où $Q_{ij} = \sum_k e_{ki} e_{kj} \left(f_k - f_k^{eq}\right)$ et $\Delta$ la largeur du filtre, égale à l'unité d'espace.

De même pour le champ thermique, le temps de relaxation est ajusté en utilisant la nouvelle diffusivité thermique :

$$\alpha_{tot} = \frac{\tau_{\alpha\text{-tot}} - 0,5}{2} = \alpha + \alpha_t = \alpha + \frac{\nu_t}{Pr_t} \quad (5)$$

où $Pr_t$ est le nombre de Prandtl turbulent, habituellement pris entre 0.3 et 1.

## 2.3. PARAMETRES DE DIFFUSION VARIABLES

Le jet plasma d'argon-azote est un écoulement à hautes températures. De ce fait, toutes les quantités physiques (viscosité, diffusivité…) varient largement en fonction de la température. Un des objectifs de l'étude présentée est d'étendre l'emploi de la méthode LBM à l'utilisation de paramètres de diffusion variables. L'idée est induite par l'analyse dimensionnelle en effet dans des cas généraux, on obtient les mêmes valeurs adimensionnelles sont obtenues pour une grandeur dans l'espace *LB* et dans l'espace *Ph* (physique):

$$\frac{\varphi_{LB}}{LB_{échelle}} = \frac{\varphi_{Ph}}{Ph_{échelle}} \qquad (6)$$

## 2.4. DOMAINE D'ETUDE ET PROFILS D'ENTREE

Le domaine d'étude est un demi-plan (tracé sur la Figure 1) subdivisé par maillage uniforme en 200x96 unités de réseau. Dans le modèle présenté, le bord OA est régi par les profils d'entrée suivants:

$$\begin{cases} u_{en}(y)/U_{max} = 1 - (y/R)^2 \\ (T_{en}(y) - T_{min})/(T_{max} - T_{min}) = 1 - (y/R)^3 \end{cases} \qquad (7)$$

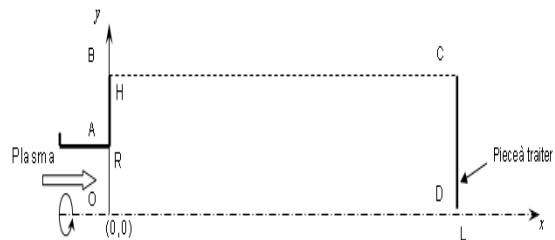

**Figure 1 - Schéma du domaine étudié**

## 4. RESULTATS

Quand l'obstacle constitué par le substrat n'est pas considéré dans le modèle du jet plasma, les profils des champs calculés diffèrent de la réalité. L'objet (substrat) constitue une frontière fixe (mur) ce qui modifie la condition classique de jet libre. Dans cette étude sont examinés les deux cas avec/sans substrat.

### 4.1. VALIDATION

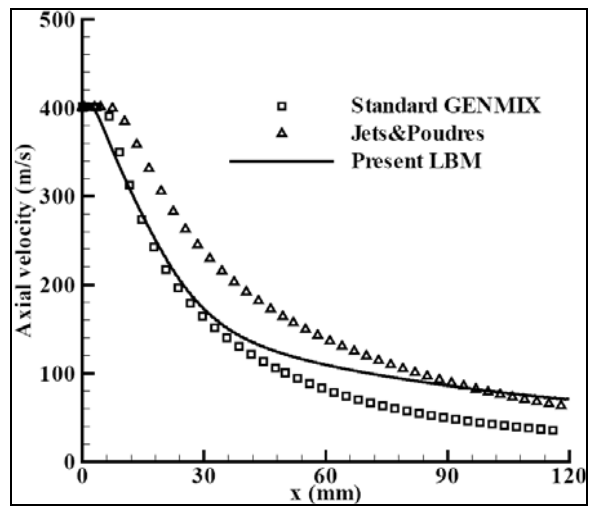

**Figure 2 - Distribution centrale de la température simulée par le modèle LBGK-LES (D2Q9-D2Q4) considérant $C_{smag}$=0.18 et $Pr_t$=0.3 en comparaison aux résultats de référence.**

La plupart des efforts numériques disponibles sont fondés sur la considération de jet libre, tels que pour les codes GENMIX et Jets&poudres. Les codes GENMIX et Jets&poudres sont considérablement améliorés et validés par application aux jets plasma [13]. Les résultats du modèle LB présenté sont comparés aux résultats de GENMIX et Jets&Poudres [14] pour les distributions centrales de la température et la composante axiale de la vitesse (voir les Figure 2 et Figure 3). Une bonne concordance est remarquée. Il est à noter que le gradient de température axiale près de l'admission (intervalle 0-20 millimètres) est proche de 140 K/mm (contre 195 K/mm et 167 K/mm pour les résultats de

GENMIX et Jets&Poudres respectivement) et le gradient de vitesse est proche de 8.4(m/s)/mm (contre 9 (m/s)/mm et 4.6 (m/s)/mm les résultats de GENMIX et Jets&Poudres respectivement) ce qui est conformes aux observations expérimentales des gradients indiquant 200K/mm et de 10m/s/mm.

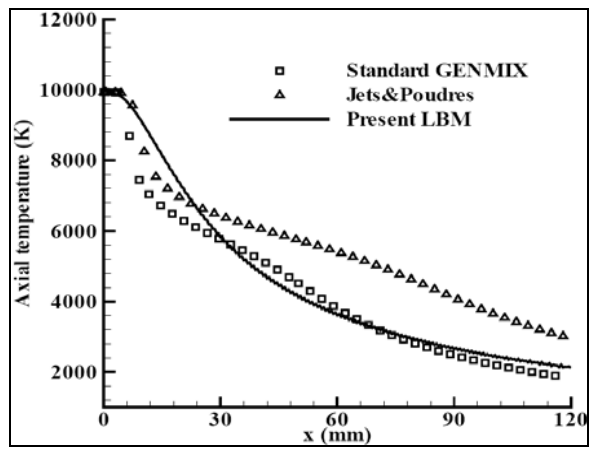

**Figure 3 -Distribution centrale de la vitesse axiale simulée par le modèle LBGK-LES (D2Q9-D2Q4) considérant $C_{smag}$=0.18 et $Pr_t$=0.3 en comparaison aux résultats de référence.**

Les figures Figure 4 et Figure 5 comparent les isothermes et la composante axiale de la vitesse aux résultats de Jets&Poudres. Les distributions calculées s'avèrent étroites en comparaison de celles de Jets&Poudres. Cependant, les expériences et d'autres travaux montrent que l'épaisseur du jet plasma (à mi-longueur du jet) est au plus environ 15mm, ce qui est bien démontré par les résultats actuels. Le profil radial gaussien est également vérifié (voir le Figure 6) pour la distribution radiale de la température à différentes sections transversales. En augmentant la vitesse maximale d'entrée, les valeurs de ligne centrale augmentent, de sorte que les franges de vitesse se déplacent en aval.

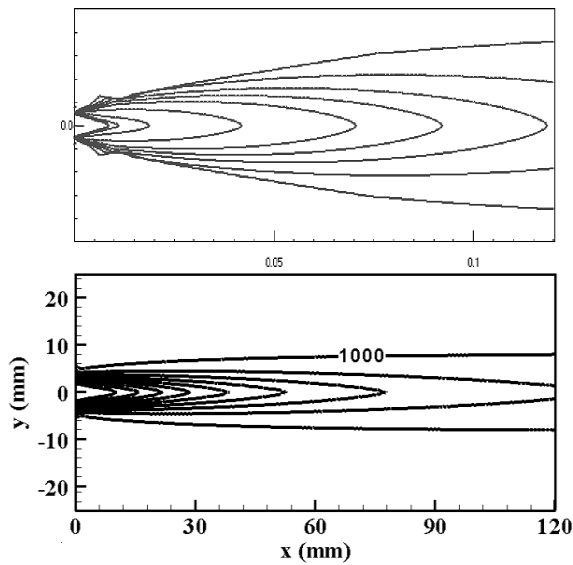

**Figure 4 -Tracés des isothermes pour le code Jets&Poudres (haut) et LBGK-LES (bas) avec 1000 K pour la ligne extérieure et l'intervalle.**

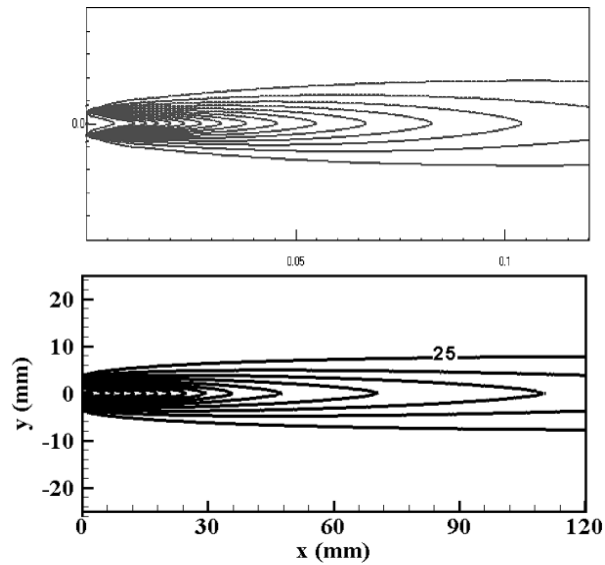

**Figure 5 -Tracés des iso-vitesse axiale pour le code Jets&Poudres (haut) et LBGK-LES (bas) avec 40 m/s pour intervalle.**

### 4.2. CAS AVEC SUBSTRAT

La normale de la pièce à traiter peut avoir plusieurs inclinaisons par rapport à la direction du jet. Nous considérons ici une incidence normale sur l'objet. La température et la vitesse maximales à l'entrée sont choisies 10000 K et 500 m/s. La cible est à 120 millimètres de la sortie de la torche. Les résultats sont présentés sur les Figure 8 et Figure 9. Il est bien démontré que les distributions de la température et de la vitesse axiale changent considérablement. Les résultats actuels sont conformes aux résultats numériques donnés par [4]. Il est, donc, plus intuitif de prendre en compte la condition aux limites due au substrat, tant que les caractéristiques du jet plasma affectent directement les histoires thermiques et dynamiques des particules injectées.

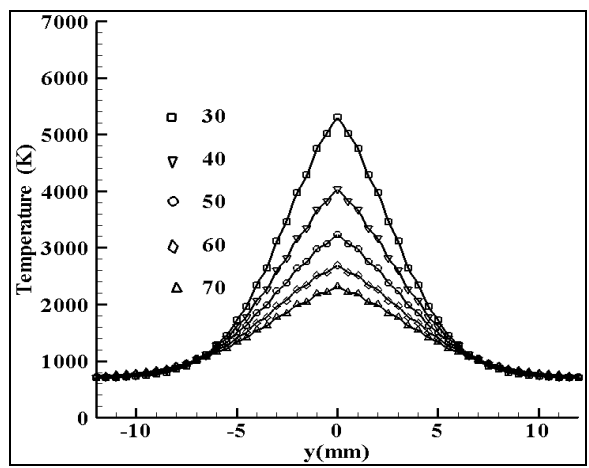

**Figure 6- Distribution radiale de la température pour différentes sections transversales simulées par le modèle LBGK-LES (D2Q9-D2Q4)**

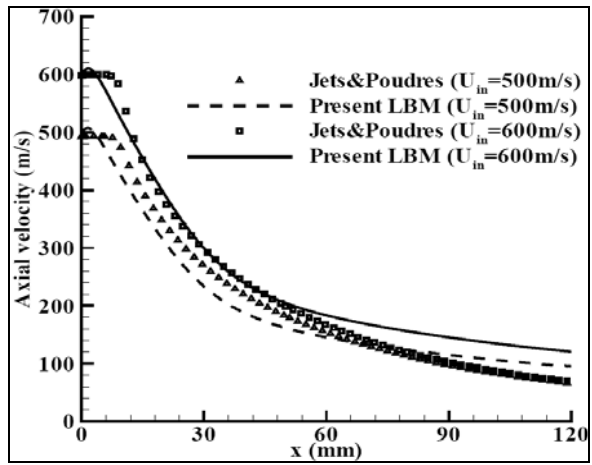

**Figure 7 -** Effet du maximum de la vitesse d'entrée sur le profil de la vitesse centrale.

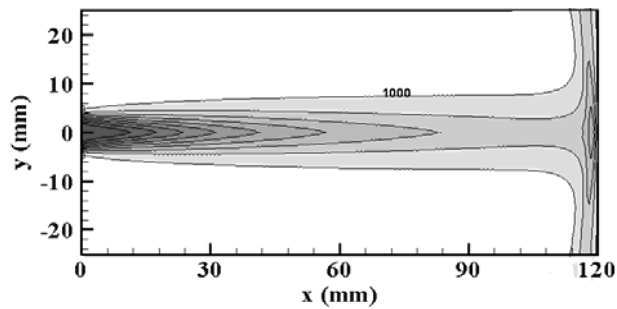

**Figure 8 -** Distribution de la température simulée par le modèle LBGK-LES pour un jet incident normalement sur le substrat, avec intervalle de 1000K.

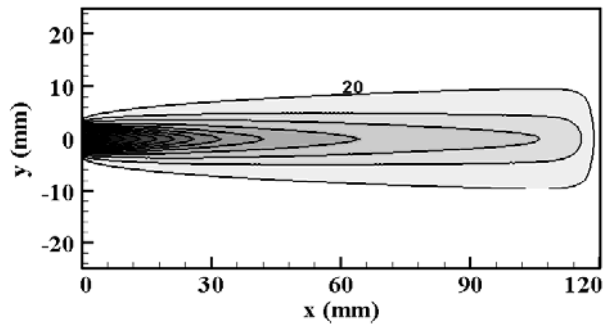

**Figure 9-** Distribution de la vitesse axiale simulée par le modèle LBGK-LES pour un jet incident normalement sur le substrat, avec intervalle de 40/s.

## 3. CONCLUSION

Dans l'étude présentée, un jet axisymétrique de plasma d'argon-azote en écoulement dans l'argon-azote est simulé en employant la méthode de Boltzmann sur réseau. Les profils calculés de température et de la composante axiale de la vitesse le long de la ligne centrale sont conformes aux résultats numériques obtenus par des codes fondés sur différents modèles de turbulence. Les distributions de la température et la vitesse axiales semblent plus représentatives du jet plasma par la méthode de Boltzmann que pour les codes GENMIX et Jets&Poudres. Compte tenu que la pièce à traiter en tant que frontière-mur affecte sensiblement la structure d'écoulement.